\documentclass[aps,prd,reprint,showpacs,longbibliography,superscriptaddress, groupedaddres,
titlepage,nofootinbib]{revtex4-1} 

\usepackage{amsthm}
\usepackage{amssymb}   
\usepackage{mathtools} 
\usepackage{hyperref}
\hypersetup{colorlinks=true,linkcolor=blue,urlcolor=blue,citecolor=blue}
\usepackage{accents}
\usepackage{tensor}
\usepackage[cal=boondox]{mathalfa}
\usepackage{lipsum}

\usepackage{color}

\usepackage[T1]{fontenc}

\begin{document}

\title{Manifestly Lorentz-covariant variables for the phase space of general relativity}

\author{Merced Montesinos} 
\email[]{merced@fis.cinvestav.mx}
\affiliation{Departamento de F{\'{\i}}sica, Cinvestav, Avenida Instituto Polit{\'e}cnico Nacional 2508, San Pedro Zacatenco, 07360 Gustavo A. Madero, Ciudad de M{\'e}xico, M{\'e}xico}

\author{Jorge Romero}
\email[]{ljromero@fis.cinvestav.mx}
\affiliation{Departamento de F{\'{\i}}sica, Cinvestav, Avenida Instituto Polit{\'e}cnico Nacional 2508, San Pedro Zacatenco, 07360 Gustavo A. Madero, Ciudad de M{\'e}xico, M{\'e}xico}

\author{Mariano Celada}
\email[]{mcelada@fis.cinvestav.mx}
\affiliation{Departamento de F{\'{\i}}sica, Cinvestav, Avenida Instituto Polit{\'e}cnico Nacional 2508, San Pedro Zacatenco, 07360 Gustavo A. Madero, Ciudad de M{\'e}xico, M{\'e}xico}
\affiliation{Departamento de F\'isica, Universidad Aut\'onoma Metropolitana Iztapalapa, San Rafael Atlixco 186, 09340 Ciudad de M\'exico, M\'exico}

\date{\today}

\begin{abstract}
	
	We present a manifestly Lorentz-covariant description of the phase space of general relativity with the Immirzi parameter. This formulation emerges after solving the second-class constraints arising in the canonical analysis of the Holst action. We show that the new canonical variables give rise to other Lorentz-covariant parametrizations of the phase space via canonical transformations. The resulting form of the first-class constraints in terms of new variables is given. In the time gauge, these variables and the constraints become those found by Barbero.

\end{abstract}

\maketitle

\section{Introduction} 

Loop quantum gravity~\cite{rovelli2004quantum,AshLewcqg2115,thiemann2007modern,RovelliLR,*Rovellicag2815} is one of the most promising candidates to successfully achieve a nonpertubative and background-independent quantization of the gravitational field. It is based on the Ashtekar-Barbero variables for general relativity~\cite{Barbero}, which emerge from the canonical analysis--with a partial gauge fixing--of the Holst action~\cite{Holst}. Although this action gives rise to the same gravitational dynamics dictated by Einstein's equations, it contains a free parameter\footnote{This parameter does not modify the classical dynamics (on shell), but becomes relevant as long as we work off shell. See for instance~\cite{RefGSGR}.} that turns out to be meaningful at the quantum level, since it shows up in the spectra of quantum observables~\cite{RoveSmolinNucP442,*AshtLewacqg14} and in the black hole entropy~\cite{RovPRLt.77.3288,*AshBaezcqg105,*Meisscqg21,*Agulloetcprl100,*EngleNuiPRL105}. It is believed that the presence of the so-called  Immirzi parameter~\cite{Immirzicqg1410} may be due to the fact that the Ashtekar-Barbero variables are obtained through the use of the time gauge (a different gauge fixing has been recently studied in~\cite{Liu_Noui}), which breaks the Lorentz invariance down to rotational invariance in order to simplify the construction of the associated quantum theory. Because of this, there have been several attempts to construct a Lorentz-covariant canonical description of the phase space of general relativity seeking to resolve the Immirzi ambiguity~\cite{Alexcqg1720,*AlexVassprd644,*AlexLivprd674,Cianfraniprl,NouiSIGMA72011,*Nouiprd84044002} (see also~\cite{Nouiprd9110} for a lower-dimensional model).

The Lorentz-covariant canonical analysis of general relativity features the presence of second-class constraints. They can be equivalently dealt with either by introducing the Dirac bracket~\cite{Alexcqg1720,*AlexVassprd644,*AlexLivprd674} or by solving them in an explicit manner~\cite{Barros,NouiSIGMA72011,*Nouiprd84044002,CelMontesRom}. In this paper we follow the latter direction. It is worth mentioning that, although the approach derived in~\cite{Barros,CelMontesRom} certainly is Lorentz covariant, it is not manifestly Lorentz covariant, since it splits the Lorentz group into boosts and rotations. Given that we would like to maintain untouched the classical symmetries of general relativity as much as possible, we ask whether it is plausible to solve the second-class constraints in a manifestly Lorentz-covariant fashion. The answer given in this paper is in the affirmative, showing that we can describe the phase space of general relativity using several canonical pairs, one of which is made up of Lorentz vectors. It turns out that the different canonical pairs are related to one another by canonical transformations that can be regarded as the Lorentz-covariant generalization of Barbero's canonical transformation. As expected, the canonical variables found in this paper lead to the Ashtekar-Barbero variables in the time gauge. For that reason, the new canonical variables reported here certainly constitute a Lorentz-covariant extension of the Ashtekar-Barbero variables.

\section{Hamiltonian action}
Our notation is as follows. Internal (Lorentz) indices are denoted by $I,J,\dots$, and take the values $\{0,i\}$, where $i=1,2,3$.  Likewise, $a,b,\dots$ label spatial coordinates. The internal indices are raised and lowered with the metric $\eta_{IJ}=\rm{diag}(\sigma,1,1,1)$, where $\sigma=-1\ (=+1)$ for Lorentzian (Euclidean) signature. The internal group, denoted by SO$(\sigma)$, corresponds to the Lorentz group SO$(1,3)$ for $\sigma=-1$ or the rotation group SO(4) for $\sigma=+1$. The weight of a tensor is sometimes indicated with the presence of a tilde over or below it, whereas the time derivative is represented by a dot over the corresponding variable. The internal tensor $\epsilon_{IJKL}$ is totally antisymmetric and such that $\epsilon_{0123}=+1$. Similarly, the spatial tensor density $\underaccent{\tilde}{\eta}_{abc}$ ($\tilde{\eta}^{abc}$) is totally antisymmetric and satisfies $\underaccent{\tilde}{\eta}_{123}=1$ ($\tilde{\eta}^{123}=+1$). The symmetrizer and antisymmetrizer are defined correspondingly by $A_{(\alpha\beta)}:=(A_{\alpha\beta}+A_{\beta\alpha})/2$ and $A_{[\alpha\beta]}:=(A_{\alpha\beta}-A_{\beta\alpha})/2$. Furthermore, for any antisymmetric quantity $A^{IJ}$ we define its internal (Hodge) dual as $\star A^{IJ}:=(1/2)\epsilon^{IJ}{}_{KL}A^{KL}$ and also the corresponding $\gamma$-valued quantity $\stackrel{(\gamma)}{A} {}^{IJ}:=P^{IJ}{}_{KL}A^{KL}=A^{IJ}+(1/\gamma)\star A^{IJ}$, where $\gamma$ is the Immirzi parameter. 

In the first-order formalism, general relativity with the Immirzi parameter can be described either by the Holst action or by a $BF$-type action supplemented with constraints~\cite{CMPR2001,*cqgrevBF} (the $BF$-type formulations play a fundamental role in the spin-foam approach to quantum gravity~\cite{alexandrov2012spin,*perez2013}). By performing the $3+1$ decomposition of the Holst action in an SO($\sigma$)-covariant fashion~\cite{Barros} (we assume that the spacetime has a topology $\mathbb{R}\times \Omega$, with $\Omega$ a spacelike three-dimensional manifold without a boundary), it takes the simplified form
\begin{equation}\label{Act3+1}
	S=\int_\mathbb{R} dt\int_\Omega d^3x\left(\stackrel{(\gamma)}{\tilde{\Pi}} {}^{aIJ}\dot{\omega}_{aIJ}-\tilde{H}\right),
\end{equation}
where $(\omega_{aIJ},\stackrel{(\gamma)}{\tilde{\Pi}}{}^{aIJ})$ are canonical coordinates\footnote{We can also use the canonical variables $(\stackrel{(\gamma)}{\omega}{}_{aIJ},\tilde{\Pi}^{aIJ})$.} and $\tilde{H}$ is the Hamiltonian (density), which is given by
\begin{equation}\label{Hamact}
\tilde{H}=\underaccent{\tilde}{N}\tilde{\tilde{\mathcal{H}}} + N^a\tilde{\mathcal{V}}_a + \xi_{IJ}\tilde{\mathcal{G}}^{IJ}+\underaccent{\tilde}{\varphi}{}_{ab}\tilde{\tilde{\Phi}}{}^{ab}+\psi_{ab}\Psi^{ab}.
\end{equation}
Here $\underaccent{\tilde}{N},N^a,\xi_{IJ},\underaccent{\tilde}{\varphi}{}_{ab}$, and $\psi_{ab}$ (of weight -2, so that $\Psi^{ab}$ has weight +3) are Lagrange multipliers imposing the constraints
\begin{subequations}
\begin{eqnarray}
&&\tilde{\mathcal{G}}^{IJ}:= D_a \stackrel{(\gamma)}{\tilde{\Pi}}\!\!{}^{aIJ} \approx 0, \label{Gauss}\\
&&\tilde{\mathcal{V}}_a := \frac{1}{2}\tilde{\Pi}^{bIJ}\stackrel{(\gamma)}{F}\!\!{}{_{baIJ}} \approx 0,\label{Vector} \\
&&\tilde{\tilde{\mathcal{H}}}:=\frac{1}{2} \tilde{\Pi}^{aIK}\tilde{\Pi}^b{}_K{}^J\stackrel{(\gamma)}{F}\!\!{}{_{abIJ}} + \sigma\Lambda g \approx 0 ,\label{Scalar}\\
&&\tilde{\tilde{\Phi}}^{ab} := -2\sigma\star\tilde{\Pi}^a{}_{IJ}\tilde{\Pi}^{bIJ} \approx 0, \label{phi}\\
&&\Psi^{ab} := \epsilon_{IJKL}\tilde{\Pi}^{(a|IM}\tilde{\Pi}^c{}_M{}^JD_c \tilde{\Pi}^{|b)KL}\approx 0\label{psi},   	
\end{eqnarray}
\end{subequations}
where $F_{abIJ}:=2\left(\partial_{[a}\omega_{b]IJ}+\omega_{[a|IK}\omega_{|b]}{}^K{}_J\right)$ is the curvature of $\omega_{aIJ}$, $D_a$ is the spatial component of the SO$(\sigma)$-covariant derivative, $g:=\det(g_{ab})$ is the determinant of the spatial metric $g_{ab}$ (the induced metric on $\Omega$) whose inverse is defined by $gg^{ab}:=(\sigma/2)\tilde{\Pi}^{aIJ}\tilde{\Pi}^{b}{}_{IJ}$, and $\Lambda$ is the cosmological constant. As a result, the constraints $\tilde{\mathcal{G}}^{IJ}$, $\tilde{\mathcal{V}}_a$, and 
$\tilde{\tilde{\mathcal{H}}}$ (they are called Gauss, vector, and scalar constraints, respectively) are first class, and generate the gauge symmetries of the theory. Sometimes, instead of $\tilde{\mathcal{V}}_a$, we consider the diffeomorphism constraint $\tilde{\mathcal{D}}_a:=\tilde{\mathcal{V}}_a+(1/2)\omega_{aIJ}\tilde{\mathcal{G}}^{IJ}$. On the other hand, the constraints $\tilde{\tilde{\Phi}}^{ab}$ and $\Psi^{ab}$ are second class. These constraints arise from the implementation of the Dirac procedure and are necessary to obtain the correct physical degrees of freedom of general relativity. It is worth realizing that, although the expressions for the constraints (\ref{Scalar}) and (\ref{psi}) differ from the corresponding ones reported in~\cite{CelMontcqg2920}, they are actually equivalent to them, since we have the relation $\tilde{\eta}^{abc}g_{cd}\tilde{\Pi}^d{}_{IJ}=\pm \sigma\epsilon_{IJKL}\tilde{\Pi}^{aKM}\tilde{\Pi}^b{}_M{}^L$ when the constraint (\ref{phi}) holds, the sign having to do with the sign ambiguity in the solution of the simplicity constraint in the $BF$ formalism.

\section{Solution of the second-class constraints}
We now solve the second-class constraints, and we proceed in such a way that we keep the explicit SO($\sigma$) covariance of the theory. The constraint (\ref{phi}) amounts to a total of six restrictions on the 18 variables $\tilde{\Pi}^{aIJ}$. Its solution is then given in terms of 12 independent variables $\tilde{B}^{aI}$ as (see~\cite{ashtekar1991lectures,peld1994115})
\begin{equation}\label{sol2nd}
	\tilde{\Pi}^{aIJ}=\epsilon\tilde{B}^{a[I}m^{J]},
\end{equation}
where $\epsilon=\pm1$ (since the constraint is quadratic in $\tilde{\Pi}^{aIJ}$) and
\begin{equation}\label{intm}
	m_I:=\frac{1}{6\sqrt{h}}\epsilon_{IJKL}\underaccent{\tilde}{\eta}_{abc}\tilde{B}^{aJ}\tilde{B}^{bK}\tilde{B}^{cL}
\end{equation}
for $h:=\det(h^{ab})$, with $h^{ab}:=\tilde{B}^{aI}\tilde{B}^b{}_{I}$. Notice that $m^I$ satisfies the identities $m_Im^I=\sigma$ and $\tilde{B}^{aI}m_I=0$. Moreover, we have the relation $4gg^{ab}=h^{ab}$ (thus, $h^{ab}$ can be regarded as the densitized metric). From now on, the inverse of $h^{ab}$, of weight -2, is denoted $h_{ab}$. This then implies the important relation $q^I{}_J:=h_{ab}\tilde{B}^{aI}\tilde{B}^b{}_J=\delta^I_J-\sigma m^Im_J$, which embodies the projector on the orthogonal plane to $m^I$.

It remains to solve the constraint (\ref{psi}). This constraint imposes six restrictions on the 18 components $\omega_{aIJ}$, meaning that the general solution of (\ref{psi}) takes the form 
\begin{equation}
	\label{wMN}
	\omega_{aIJ}= M_a{}^b{}_{IJK}C_{b}{}^{K} + N_{aIJ},
\end{equation}
where the first and the second terms on the right are the homogeneous and particular solutions of (\ref{psi}), respectively. Here, the 12 variables $C_{aI}$ parametrize the homogeneous solution. To determine $M_a{}^b{}_{IJK}$ and $N_{aIJ}$, we demand the independent variables $\tilde{B}^{aI}$ and  $C_{aI}$ to be canonically conjugate to each other. This is allowed, since the solution \eqref{sol2nd} induces a reduction of the symplectic structure in \eqref{Act3+1},
\begin{equation}
	\label{ss}
	\stackrel{(\gamma)}{\tilde{\Pi}} {}^{aIJ}\dot{\omega}_{aIJ}=\tilde{B}^{aI}\dot{C}_{aI},
\end{equation}
where we have defined
\begin{equation}
	\label{C}
	C_{aI} := \epsilon \left( \stackrel{(\gamma)}{\omega}\!\!{}_{aIJ}m^J + m_I \stackrel{(\gamma)}{\omega}\!\!{}_{bJK} h_{ac} \tilde{B}^{cJ}\tilde{B}^{bK}\right).
\end{equation} 
Solving jointly \eqref{psi} and \eqref{C}, we obtain 
\begin{eqnarray}
	 M_a{}^b{}_{IJK} &=& \epsilon \sigma \left[ - \delta_{a}^{b} m_{[I} \eta_{J] K} + \delta_{a}^{b} \left( P^{-1}\right)_{I J K L} m^{L} \phantom{\dfrac{1}{2}}\right. \nonumber \\
	 & & - \left( P^{-1}\right)_{I J L M} h_{a c} \tilde{B}^{cL} \tilde{B}^{bM} m_{K}  \nonumber \\
	 & &  \left. +   \dfrac{1}{\gamma}\star \left( P^{-1}\right)_{IJLM} h_{a c} \tilde{B}^{bL}m^{M} \tilde{B}^{c}{}_{K} \right]\label{Mom}, \\
	 \label{N}
	   N_{aIJ} & = & \underaccent{\tilde}{ \lambda}_{ab} \left( -\sigma \epsilon_{IJKL}\tilde{B}^{bK}m^L +\dfrac{2}{\gamma} \tilde{B}^{b}{}_{[I}m_{J]} \right)\label{Nom},
\end{eqnarray}
with $\left( P^{-1}\right)_{IJ}{}^{KL}$ being the inverse of $P^{IJ}{}_{KL}$ that satisfies $ P^{IJ}{}_{KL}(P^{-1})^{KL}{}_{MN}=\delta^I_{[M}\delta^J_{N]}$ and
\begin{equation}
	\underaccent{\tilde}{\lambda}_{ab} := \dfrac{\sigma}{2} \epsilon_{IJKL}\left( h_{ab}h_{cd} - 2 h_{c(a}h_{b)d}\right)\tilde{B}^{cI}\tilde{B}^{fJ}m^{L}\partial_f \tilde{B}^{dK}.
\end{equation}
In short, the expression \eqref{wMN}, together with \eqref{Mom} and \eqref{Nom}, is the solution of (\ref{psi}). Notice that the quantities $N_{aIJ}$ (or $\underaccent{\tilde}{ \lambda}_{ab}=\underaccent{\tilde}{ \lambda}_{ba}$), which can be thought of as the components of the connection not showing up in the symplectic structure \eqref{ss}, are the ones getting fixed by the solution of (\ref{psi}).

Up to now, we have gotten rid of the second-class constraints, leaving in the process a phase space parametrized by the canonical pair $(C_{aI},\tilde{B}^{aI})$ subject to first-class constraints only. It is then necessary to rewrite these constraints in terms of the new canonical variables.  To carry out this, let us first introduce the covariant derivative compatible with $\tilde{B}^{aI}$ satisfying
\begin{equation}\label{covder}
\nabla_a \tilde{B}^{bI}:=\partial_a\tilde{B}^{bI}+\Gamma_a{}^I{}_J\tilde{B}^{bJ}+\Gamma^b{}_{ac}\tilde{B}^{cI}-\Gamma^c{}_{ac}\tilde{B}^{bI}=0,
\end{equation}
where $\Gamma_{aIJ}$ ($=-\Gamma_{aJI}$) takes values in the Lie algebra of SO($\sigma$) and $\Gamma^b{}_{ac}=\Gamma^b{}_{ca}$. These 36 equations allow us to completely fix the 36 quantities $\Gamma_{aIJ}$ and $\Gamma^b{}_{ac}$. In fact, $\Gamma^b{}_{ac}$ is nothing but the Christoffel symbol for the metric $g_{ab}$, that is, the Levi-Civita connection compatible with the spatial metric.

The introduction of $\Gamma^{a}{}_{bc}$ and $\Gamma_{aIJ}$, together with the solutions for $\tilde{\Pi}^{aIJ}$ and $\omega_{aIJ}$, allows us to express the  first-class constraints (\ref{Gauss})-(\ref{Scalar}) as
\begin{subequations}\label{constr2}
	\begin{eqnarray}
	&&\hspace{-1.5mm}\tilde{\mathcal{G}}^{IJ}= \tilde{B}^{a[I}C_a{}^{J]}+2\epsilon P^{IJ}{}_{KL}\tilde{B}^{a[M}m^{K]}\Gamma_a{}^L{}_M\approx 0, \label{GaussC}\\
	&&\hspace{-1.5mm}\tilde{\mathcal{V}}_a = \nabla_{[b}\left(\tilde{B}^{bI}C_{a]I}\right)+\epsilon \tilde{B}^{b[I}m^{K]}\stackrel{(\gamma)}{\Gamma}_{aIJ}\Gamma_b{}^J{}_K\nonumber\\
	&&\hspace{7.5mm}-\epsilon\sigma\tilde{\mathcal{G}}^{IJ}\Bigl(C_{aI}-\epsilon \stackrel{(\gamma)}{\Gamma}_{aIK}m^K\Bigr)m_J\approx 0,\label{VectorC}\\
	&&\hspace{-1.5mm}\tilde{\tilde{\mathcal{H}}}= -\frac{\sigma}{8}\tilde{B}^{aI}\tilde{B}^{bJ}R_{abIJ} +\frac{1}{4}\tilde{B}^{a[I|}\tilde{B}^{b|J]}\Biggl[C_{aI}C_{bJ}\nonumber\\
	&&-2\epsilon C_{aI}\stackrel{(\gamma)}{\Gamma}_{bJK}m^K+\left(\Gamma_{aIL}+\frac{2}{\gamma}\star\Gamma_{aIL}\right)\Gamma_{bJK}m^Km^L\nonumber \\
	&&+\frac{1}{\gamma^2}q^{KL}\Gamma_{aIK}\Gamma_{bJL}\Biggr]-\frac{\epsilon}{2}\tilde{B}^{aI} m^J\nabla_a\tilde{\mathcal{G}}_{IJ} +\frac{\sigma\Lambda}{8} \sqrt{h}\approx 0,\nonumber\\
	\label{ScalarC}
	\end{eqnarray}
\end{subequations}
where $R_{abIJ}$ is the curvature of $\Gamma_{aIJ}$ and the terms proportional to $\tilde{\mathcal{G}}^{IJ}$ squared have been dropped. It is important to note that although the constraints look rather complicated, they collapse to the ones of the Ashtekar-Barbero's formulation once we take the time gauge (see below), meaning that the variables $(C_{aI},\tilde{B}^{aI})$ may be regarded as the explicitly SO($\sigma$)-covariant version of the Ashtekar-Barbero variables. Nonetheless, at the fully covariant level, we can explore different parametrizations of the phase space of the theory.

\section{Other Lorentz-covariant parametrizations of the phase space}

Let us consider a change of coordinates in which the momentum variables $\tilde{B}^{aI}$ remain unchanged, while the configuration variables take the form 
\begin{subequations}\label{transfcan}
\begin{eqnarray}
	\hspace{-7mm}C_{aI}=&&\epsilon\left(\Gamma_{aIJ}m^J+h_{ab}\tilde{B}^{bJ}\tilde{B}^{cK}\Gamma_{cJK}m_I\right)+K_{aI}\label{transfcan1}\\
	=&&\epsilon\left(\stackrel{(\gamma)}{\Gamma}_{aIJ}m^J\hspace{-.5mm}+h_{ab}\tilde{B}^{bJ}\tilde{B}^{cK}\stackrel{(\gamma)}{\Gamma}_{cJK}m_I\hspace{-.5mm}\right)\hspace{-.5mm}+Q_{aI}.\label{transfcan2}
\end{eqnarray}
\end{subequations}
These transformations give rise to the canonical pairs $(K_{aI}, \tilde{B}^{aI})$ and $(Q_{aI}, \tilde{B}^{aI})$. Indeed,  a direct substitution of \eqref{transfcan} in \eqref{ss} results in
 \begin{subequations}\label{canvar}
\begin{eqnarray}
\tilde{B}^{aI}\dot{C}_{aI}\label{canvar1}	&=& \partial_a\left(\epsilon\dot{\tilde{B}}{}^{aI}m_I\right)+\tilde{B}^{aI}\dot{K}_{aI}\label{canvar2}\\
	&=&\partial_a\left(\epsilon\dot{\tilde{B}}{}^{aI}m_I-\frac{\epsilon\sigma}{2\gamma}\sqrt{h}\tilde{\eta}^{abc}h_{bd}h_{ce}\dot{\tilde{B}}^{dI}\tilde{B}^e{}_I\right)\nonumber\\
	& &+\tilde{B}^{aI}\dot{Q}_{aI}.\label{canvar3}
\end{eqnarray}
\end{subequations}
We see that, in any case, the symplectic structures in \eqref{canvar} differ from one another by a divergence (which does not contribute if the spatial manifold has no boundary or if suitable boundary conditions are imposed when it has a boundary); this shows that the transformations associated to (\ref{transfcan}) are canonical. Alternatively, given that the pair $(C_{aI}, \tilde{B}^{aI})$ is canonical, it can be shown by using the Poisson brackets that the relations (\ref{transfcan}) induce canonical transformations among the different pairs considered. Actually, the only complicated bracket is the one involving the new configuration variable with itself, but, by following a procedure close to that of~\cite{ashtekar1991lectures,thiemann2007modern}, it can be shown that it vanishes because the terms between parentheses in (\ref{transfcan}) can be derived from a potential in each case.

For the sake of completeness, we display in the following lines the form of the constraints in the new sets of canonical variables. In the canonical coordinates $(K_{aI},\tilde{B}^{aI})$, the constraints read
\begin{subequations}\label{constr3}
	\begin{eqnarray}
	&&\hspace{-3mm}\tilde{\mathcal{G}}^{IJ}= \tilde{B}^{a[I}K_a{}^{J]}+\frac{\epsilon}{\gamma} \epsilon^{IJ}{}_{KL}\tilde{B}^{a[M}m^{K]}\Gamma_a{}^L{}_M\approx 0 , \label{GaussK}\\
	&&\hspace{-3mm}\tilde{\mathcal{V}}_a = \nabla_{[b}\left(\tilde{B}^{bI}K_{a]I}\right)+\frac{\epsilon}{\gamma} \tilde{B}^{b[I}m^{K]}\star\Gamma_{aIJ}\Gamma_b{}^J{}_K\nonumber\\
	&&\hspace{5mm}-\epsilon\sigma\tilde{\mathcal{G}}^{IJ}\left(K_{aI}-\frac{\epsilon}{\gamma}\star\Gamma_{aIK}m^K\right)m_J\approx 0,\label{VectorK} \\
	&&\hspace{-3mm}\tilde{\tilde{\mathcal{H}}}= -\frac{\sigma}{8}\tilde{B}^{aI}\tilde{B}^{bJ}R_{abIJ} +\frac{1}{4}\tilde{B}^{a[I|}\tilde{B}^{b|J]}\Biggl[K_{aI}K_{bJ}\nonumber\\
	&&\hspace{5mm}-\frac{2\epsilon}{\gamma} K_{aI}\star\Gamma_{bJK}m^K+\frac{1}{\gamma^2}q^{KL}\Gamma_{aIK}\Gamma_{bJL}\Biggr]\nonumber \\
	&&\hspace{5mm}-\frac{\epsilon}{2}\tilde{B}^{aI} m^J\nabla_a\tilde{\mathcal{G}}_{IJ} +\frac{\sigma\Lambda}{8} \sqrt{h}\approx 0,\label{ScalarK}
	\end{eqnarray}
\end{subequations}
whereas for the set $(Q_{aI},\tilde{B}^{aI})$, the constraints are simply
\begin{subequations}\label{constr1}
	\begin{eqnarray}
	&&\tilde{\mathcal{G}}^{IJ}= \tilde{B}^{a[I}Q_a{}^{J]} \approx 0, \label{Gauss1}\\
	&&\tilde{\mathcal{V}}_a = \nabla_{[b}\left(\tilde{B}^{bI}Q_{a]I}\right)-\epsilon\sigma\tilde{\mathcal{G}}^{IJ}Q_{aI}m_J\approx 0,\label{Vector1} \\
	&&\tilde{\tilde{\mathcal{H}}}= -\frac{\sigma}{8}\tilde{B}^{aI}\tilde{B}^{bJ}R_{abIJ} +\frac{1}{4}\tilde{B}^{a[I|}\tilde{B}^{b|J]}Q_{aI}Q_{bJ}\nonumber\\
	&&\hspace{8.5mm}-\frac{\epsilon}{2}\tilde{B}^{aI} m^J\nabla_a\tilde{\mathcal{G}}_{IJ} +\frac{\sigma\Lambda}{8} \sqrt{h} \approx 0.\label{Scalar1}
	\end{eqnarray}
\end{subequations}
As can be inferred from (\ref{Gauss1}), both $\tilde{B}^{aI}$ and $Q_{aI}$ transform as SO($\sigma$) vectors. Meanwhile, in the other cases we observe that $\tilde{B}^{aI}$ still transforms as an SO($\sigma$) vector but, according to (\ref{GaussC}) and (\ref{GaussK}), the properties of $C_{aI}$ and $K_{aI}$ under SO($\sigma$) transformations are nontrivial (but they contain a vector part). Furthermore, notice that the diffeomorphism constraint can be expressed as
\begin{equation}\label{diff}
	\tilde{\mathcal{D}}_a=\tilde{B}^{bI}\partial_{[b}U_{a]I}+\frac{1}{2}U_{aI}\partial_b\tilde{B}^{bI},
\end{equation}
for $U_{aI}=C_{aI},\ K_{aI}$ or $Q_{aI}$, which implies that these variables transform as 1-forms under spatial diffeomorphisms.

It is worth realizing that the Immirzi parameter does not explicitly appear in the constraints (\ref{constr1}) (it however affects the constants in front of some of the terms proportional to $\tilde{\mathcal{G}}^{IJ}$ squared, which we have neglected). Hence, for the canonical variables $(Q_{aI},\tilde{B}^{aI})$, the Immirzi parameter remains undetectable at the classical level. The constraints (\ref{constr1}) actually have the same form as those obtained for the case of the Palatini action (with a cosmological constant) alone (take the limit $\gamma\rightarrow\infty$; see also~\cite{peld1994115}). Therefore, the canonical transformation (\ref{transfcan}) allows us to connect the Hamiltonian form of the Holst action with that of Palatini's, reinforcing the classical equivalence of them.

\section{Time gauge}

Here we show how the previous SO($\sigma$)-covariant variables lead to the Ashtekar-Barbero ones. To that end, we adopt the time gauge, which allows us to reduce SO($\sigma$) to its compact subgroup SO(3) by fixing the boosts. In the present framework, this is accomplished by setting $\tilde{B}^{a0}=0$, which is equivalent to $m^i=0$ for a nondegenerate $\tilde{B}^{ai}$ (assumed in what follows). This gauge condition in turn implies that the boost constraint $\tilde{\mathcal{G}}_{0i}$ (in any case) must be solved at once, since $\{\tilde{B}^{a0}(x),\tilde{\mathcal{G}}^{0i}(y)\}=(\sigma/2)\tilde{B}^{ai}\delta^3(x,y)$ defines a nonsingular matrix and so $\tilde{B}^{a0}$ and $\tilde{\mathcal{G}}^{0i}$ form a second-class pair.  From \eqref{GaussC}, the solution of $\tilde{\mathcal{G}}^{0i}$ reads $C_{a0}=\sigma m^0 \tilde{B}^{bi}\partial_{b}\underaccent{\tilde}{B}_{ai}$ (for the other pairs of variables we obtain $K_{a0}=0$, $Q_{a0}=0$), with $\underaccent{\tilde}{B}_{ai}$ being the inverse of $\tilde{B}^{ai}$. In consequence, the remaining internal symmetry is SO(3), whose infinitesimal generator is the constraint $\tilde{\mathcal{G}}^{ij}$.

In the time gauge, we have, from (\ref{covder}), that $\Gamma_{a0i}=0$, whereas $\Gamma_{ai}:=(1/2)\epsilon_{ijk}\Gamma_a{}^{ij}$ ($\epsilon_{ijk}:=\epsilon_{0ijk}$) becomes the spin connection compatible with $\tilde{B}^{ai}$,
\begin{equation}\label{spinconn}
	\Gamma_{ai}=\epsilon_{ijk}\left(\partial_{[b}\underaccent{\tilde}{B}_{a]}{}^j+\underaccent{\tilde}{B}_a{}^{[l|}\tilde{B}^{c|j]}\partial_b\underaccent{\tilde}{B}_{cl}\right)\tilde{B}^{bk}.
\end{equation}
 The canonical transformation (\ref{transfcan}) then takes the form
\begin{equation}\label{tgct}
	A_{ai}=-\epsilon m^0Q_{ai}+\frac{1}{\gamma}\Gamma_{ai},
\end{equation}
with $A_{ai}:=-\epsilon m^0C_{ai}\ (=-\epsilon m^0K_{ai})$. Since $Q_{ai}$ is a vector and $\Gamma_{ai}$ is a connection, $A_{ai}$ is a connection, the gauge group being in this case SO(3). The canonical transformation (\ref{tgct}) is nothing but Barbero's canonical transformation~\cite{Barbero} (Barbero picks $\gamma=-1$ to rewrite the constraints), which here was derived from (\ref{transfcan}). Thus, the latter corresponds to the SO($\sigma$)-covariant version of the canonical transformation implemented by Barbero in order to obtain his canonical description of the phase space of general relativity. Accordingly, the quantity $-\epsilon m^0Q_{ai}$ can be related to the extrinsic curvature in the SO(3) ADM formalism~\cite{AshBalJo,ashtekar1991lectures}, while $\gamma A_{ai}$ corresponds to the Ashtekar-Barbero connection. 

Let us introduce the densitized triad $\tilde{E}^{ai}$ through $\tilde{B}^{ai}=:-2\epsilon m^0 \tilde{E}^{ai}$. The phase space is now parametrized by the pair $(A_{ai},\tilde{E}^{ai})$ satisfying the canonical commutation relation $\{A_{ai}(x),\tilde{E}^{bj}(y)\}=(1/2)\delta_a^b\delta_i^j\delta^3(x,y)$. Notice that since (\ref{spinconn}) is invariant under constant rescalings, $\Gamma_{ai}$ takes exactly the same form in terms of the densitized triad. Since $A_{ai}$ is an SO(3) connection ($\gamma A_{ai}$ to be more precise), let $F_{abi}:=2\partial_{[a}A_{b]i}-\gamma\epsilon_{ijk}A_{a}{}^j A_{b}{}^k$ be its field strength. Using (\ref{tgct}), the next identity arises right away,
\begin{equation}\label{relcurvs}
F_{abi}=-2\epsilon m^0\nabla_{[a}Q_{b]i}+\frac{1}{\gamma}R_{abi}-\gamma\epsilon_{ijk}Q_{a}{}^j Q_{b}{}^k,
\end{equation}
where $R_{abi}:=(1/2)\epsilon_{ijk}R_{ab}{}^{jk}=2\partial_{[a}\Gamma_{b]i}-\epsilon_{ijk}\Gamma_{a}{}^j \Gamma_{b}{}^k$. With the previous expression at hand, the constraints (\ref{Gauss1})-(\ref{Scalar1}) can be given, in the time gauge, the form
\begin{subequations}\label{barbero}
	\begin{eqnarray}
	&&\hspace{-4mm}\tilde{\mathcal{G}}_i:=\frac{1}{2}\epsilon_{ijk}\tilde{\mathcal{G}}^{jk}=\frac{1}{\gamma}\left(\partial_a\tilde{E}^a{_i}-\gamma\epsilon_{ijk}A_a{}^j\tilde{E}^{ak}\right) \approx 0, \label{Gauss2}\\
	&&\hspace{-4mm}\tilde{\mathcal{V}}_a = \tilde{E}^{bi}F_{bai} +\left(\gamma A_{ai}-\Gamma_{ai}\right)\tilde{\mathcal{G}}^i\approx 0,\label{Vector2} \\
	&&\hspace{-4mm}\tilde{\tilde{\mathcal{H}}}= -\frac{1}{2\gamma}\epsilon_{ijk}\tilde{E}^{ai}\tilde{E}^{bj}\left[F_{ab}{}^k+\left(\sigma\gamma-\gamma^{-1}\right)R_{ab}{}^k\right]\nonumber\\
	&&\hspace{4.5mm}+\frac{1}{\gamma} \tilde{E}^{ai}\nabla_a \tilde{\mathcal{G}}_i +\sigma\Lambda |\tilde{\tilde{E}}| \approx 0 ,\label{Scalar2}
	\end{eqnarray}
\end{subequations}
with $\tilde{\tilde{E}}:=\det\tilde{E}^{ai}$. These are precisely the constraints of the Ashtekar-Barbero formulation of general relativity. It is worth stressing that although we followed a path resembling the one walked by Barbero, the previous result can be achieved regardless of the canonical pair considered. In particular, since we may think of the constraints \eqref{constr2} as the constraints \eqref{constr1} with the canonical transformation \eqref{transfcan2} already implemented, we can obtain \eqref{barbero} directly from \eqref{constr2} in the time gauge. Notice that in our approach the Immirzi parameter plays the role of a coupling constant for the local SO(3) symmetry.

\section{Conclusions}

In this paper we have solved the second-class constraints arising in the Hamiltonian analysis of first-order general relativity with the Immirzi parameter in a manifestly SO($\sigma$)-covariant fashion (recall that $\sigma=-1$ corresponds to the Lorentz group). As a result, we obtained a description of the phase space involving only first-class constraints that exhibits a dependence on the Immirzi parameter. The associated canonical variables, which we called 
$(C_{aI},\tilde{B}^{aI})$, turn out to be related to other two sets of SO($\sigma$)-covariant variables $(K_{aI},\tilde{B}^{aI})$ and $(Q_{aI},\tilde{B}^{aI})$ by means of the canonical transformations determined by \eqref{transfcan}. In contrast to both $K_{aI}$ and $C_{aI}$, whose transformation law under local SO($\sigma$) transformations is nontrival [as can be deduced from (\ref{GaussC}) and (\ref{GaussK})], the configuration variable $Q_{aI}$ transforms as an SO($\sigma$) vector (because of the form of the canonical transformation, $\tilde{B}^{aI}$ is an internal vector in any case). In terms of the canonical pair $(Q_{aI},\tilde{B}^{aI})$, the constraints take the same form as those resulting from the canonical analysis of the Palatini action, thus eliminating the Immirzi parameter from the canonical theory. Remarkably, in the time gauge the canonical transformation \eqref{transfcan2} becomes the transformation implemented by Barbero in order to obtain his canonical formulation of general relativity~\cite{Barbero}, the spatial components $C_{ai}$ (or $K_{ai}$) being related to the Ashtekar-Barbero connection. In consequence, the canonical variables $(C_{aI},\tilde{B}^{aI})$ [or $(K_{aI},\tilde{B}^{aI})$] can be regarded as the SO($\sigma$)-covariant extension of the Ashtekar-Barbero variables.

To sum up, here we have clarified the origin of the Ashtekar-Barbero variables and their relation to the Holst action. More interesting is the Hamiltonian formulation embodied in the constraints (\ref{constr2}), (\ref{constr3}) or (\ref{constr1}), which explicitly exhibits the SO($\sigma$) invariance of the original theory (on the other hand, the Ashtekhar-Barbero formulation breaks it by resorting to the time gauge). This feature is appealing since the Lorentz symmetry is thought to be one of the fundamental symmetries of nature, and so it would be desirable to preserve it to the utmost in a quantum theory of gravity. Thus, the Hamiltonian formulation presented in this paper could be an interesting starting point for new developments to approach the quantization of gravity, something we think will provide meaningful results. In particular, we expect that this formulation might help to determine the fate of the Immirzi parameter in quantum gravity once for all.

\acknowledgements

We thank Giorgio Immirzi, Karim Noui, Alejandro Perez, Carlo Rovelli, Jos\'e David Vergara, and Jos\'e A. Zapata for their valuable comments. This work was supported in part by Consejo Nacional de Ciencia y Tecnolog\'{i}a (CONACyT), M\'{e}xico, Grant No. 237004-F. 


\bibliography{references}

\end{document}